\documentclass[conference]{IEEEtran}
\IEEEoverridecommandlockouts

\usepackage{subcaption}
\usepackage{hyperref}

\usepackage{cite}
\usepackage{amsmath,amssymb,amsfonts}
\usepackage{algorithmic}
\usepackage{graphicx}
\usepackage{textcomp}
\usepackage{xcolor}
\def\BibTeX{{\rm B\kern-.05em{\sc i\kern-.025em b}\kern-.08em
    T\kern-.1667em\lower.7ex\hbox{E}\kern-.125emX}}
\begin{document}
\title{Representing Web Applications As\\
Knowledge Graphs

\author{\IEEEauthorblockN{Yogesh Chandrasekharuni}
\IEEEauthorblockA{\textit{Skil Inc, USA} \\
yogesh@skil.ai}
}}

\maketitle

\begin{abstract}
Traditional methods for crawling and parsing web applications predominantly rely on extracting hyperlinks from initial pages and recursively following linked resources. This approach constructs a graph where nodes represent unstructured data from web pages, and edges signify transitions between them. However, these techniques are limited in capturing the dynamic and interactive behaviors inherent to modern web applications. In contrast, the proposed method models each node as a structured representation of the application’s current state, with edges reflecting user-initiated actions or transitions. This structured representation enables a more comprehensive and functional understanding of web applications, offering valuable insights for downstream tasks such as automated testing and behavior analysis.
\end{abstract}

\section{Introduction}
\label{sect:introduction}

Web applications require rich data representation for downstream tasks such as automation testing, user behavior analysis, and functional verification. Traditional web parsers operate through a structured yet simplistic algorithm:

\begin{enumerate}
    \item Initialize a queue with the starting page.
    \item Set a maximum depth (if applicable) and initialize the current depth to zero.
    \item While the queue is not empty and the maximum depth is not exceeded:
    \begin{enumerate}
        \item Dequeue the next page from the queue.
        \item If the page has not been visited:
        \begin{enumerate}
            \item Navigate to the page.
            \item Extract the desired data and store it as a node.
            \item Extract all hyperlinks from the page.
            \item Add all unseen and unvisited hyperlinks to the queue.
            \item Mark the current page as visited.
        \end{enumerate}
        \item Increment the depth if moving to a new level.
    \end{enumerate}
    \item Stop when all pages are visited or the maximum depth is reached.
\end{enumerate}

While this approach effectively scrapes static web applications, it falls short in handling dynamic applications, where significant portions of the application are unreachable through simple hyperlink navigation. Modern web applications often follow structured user flows, which involve interaction beyond hyperlinks. For instance, in an e-commerce site, reaching the checkout page might require several actions: searching for a product, adding it to the cart, entering a delivery location, and only then accessing the checkout. Traditional parsers, which rely solely on clicking hyperlinks, cannot capture such dynamic flows and are limited in their ability to represent the application’s state accurately.

Additionally, many web applications exhibit variability at the same endpoint depending on the user's context. For example, a checkout page may display "Ready to purchase" for one user and "Item cannot be delivered to your location" for another, based on the delivery address provided.

In this work, the proposed solution overcomes these limitations by representing each unique state of a web application as a node, with edges defined by specific actions taken within the application. This method captures the full complexity of user flows, allowing for a more accurate and interpretable knowledge representation of web applications.

\section{Background}
\label{sect:background}

Early web crawlers, such as \textit{World Wide Web Wanderer} (1993) \cite{1}, were primarily designed to map the size of the web by collecting basic HTML from static websites \cite{2}. As the web expanded, tools like \textit{JumpStation} emerged, becoming the first search engine to use crawlers for indexing web content \cite{3}. These early systems, however, were limited to handling static web content, as dynamic web pages driven by JavaScript and AJAX had not yet become widespread.

The emergence of dynamic content significantly complicated the process of web scraping for traditional parsers. Frameworks such as \textit{Beautiful Soup} (2004) were introduced to facilitate the extraction of structured data from increasingly complex web pages. Although effective for parsing static HTML content, these tools were inherently limited in their capacity to handle dynamic, JavaScript-driven web elements or to interact with user-initiated events. As modern web applications began to rely heavily on dynamic content loading and client-side interactions, more advanced methodologies became necessary to accurately capture these behaviors. Several tools have been developed to address these challenges. \textit{Selenium} \cite{4} is widely used for automating browser interactions, allowing developers to simulate user actions such as clicking, typing, and submitting forms.

To address these limitations, visual web scraping tools like \textit{Octoparse} \cite{5} emerged, offering user-friendly interfaces that allowed non-programmers to automate the extraction of both static and dynamic website data. These tools simulate user behavior, such as clicks and form submissions, to capture data. However, tools like Octoparse lack self-exploration capabilities and are unable to reason through or autonomously navigate complex web applications. As a result, they struggle to capture the full range of state transitions and user interactions that are essential for modeling modern, dynamic web applications.

More recent efforts have focused on combining dynamic analysis with event-based crawling techniques. For instance, \textit{jÄk} \cite{6} employs dynamic analysis to hook into JavaScript APIs and detect network events, dynamically-generated URLs, and user form submissions. By leveraging a navigation graph, jÄk can explore 86\% more of a web application's surface compared to traditional approaches. Similarly, \textit{Crawljax} \cite{7} uses state abstraction to generate a state-flow graph for AJAX-based applications, which can be used to automate testing of dynamic user flows.

Knowledge graph-based systems, such as \textit{Squirrel} \cite{8}, have been proposed to crawl the semantic web and represent data in structured formats. However, these approaches are typically limited to RDF-based web data and do not capture the dynamic, user-driven interactions seen in modern web applications.

Our proposed solution builds upon these existing tools by representing web applications as a knowledge graph. In this approach, each node represents a unique state of the application, while edges represent user actions leading to transitions between states. This structured representation allows us to capture dynamic workflows and state transitions more effectively, making the system ideal for downstream tasks like automated testing and error handling in complex web environments.
\section{Methodology}
\label{sect:methodology}

The key components of the system are designed to capture the dynamic behavior of web applications, including user interactions and state transitions. The system consists of three main components: the \textit{Functionality Inferring Module}, the \textit{Action Executor}, and the \textit{Reward/Penalty Model}. Each component interacts with the others to build an interpretable, action-based graph of the web application.

\begin{figure*}[tb]
	\centering
	\includegraphics[width=0.8\textwidth,keepaspectratio]{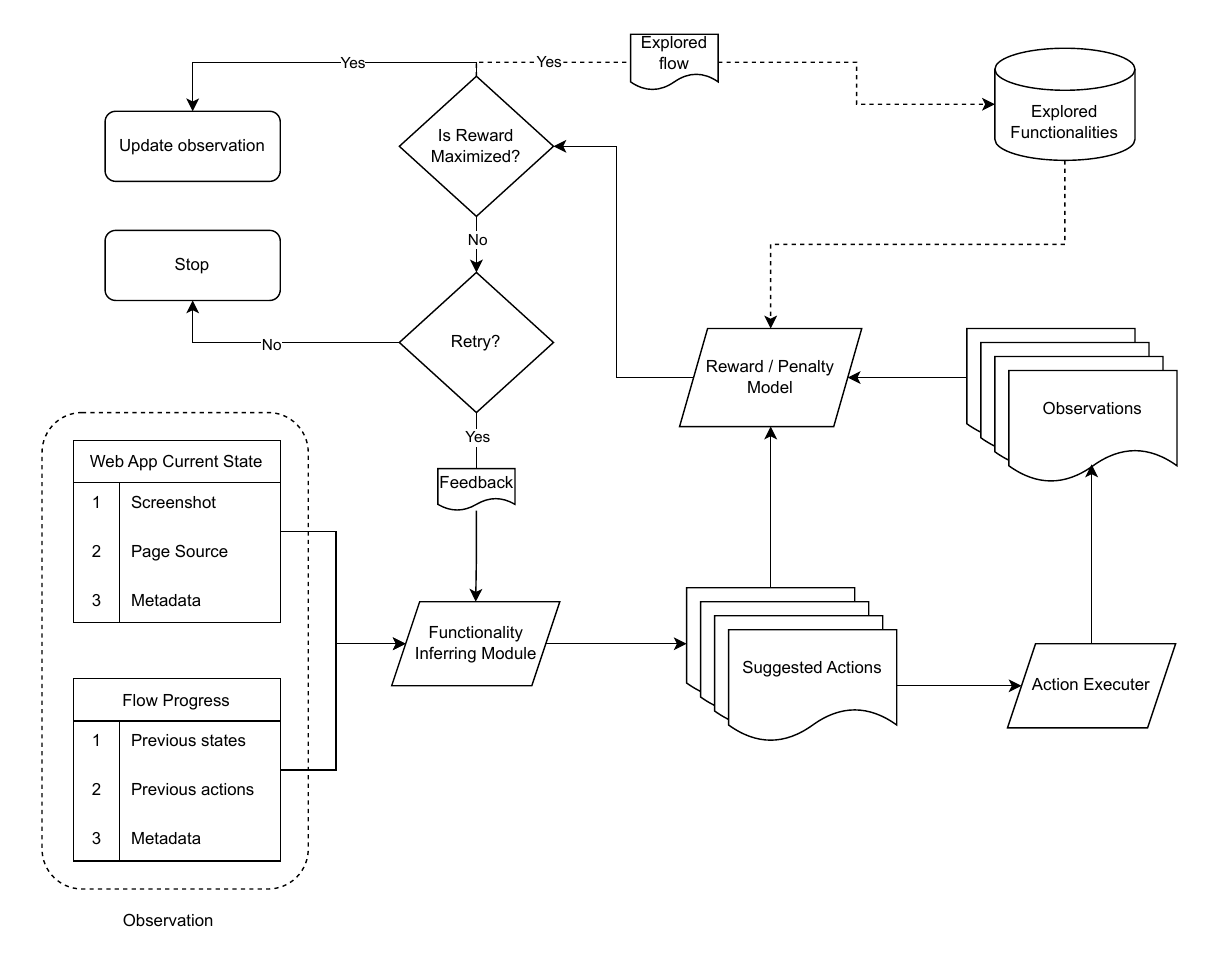}
	\caption{System Overview of the Proposed Methodology}
	\label{fig:figure-1}
\end{figure*}
\subsection{State}
\label{sect:state}

In this system, a \textit{state} refers to the configuration of a web application at a specific point in time, characterized by its visual and structural properties. Each state encapsulates the outcomes of user interactions and changes within the web application, offering a detailed snapshot of both its user interface and underlying functionality.

A state is defined by the following key components:
\begin{itemize}
    \item \textbf{Screenshot:} A visual capture of the web application's interface at a particular moment, serving as a reference for the graphical presentation as perceived by the user.
    \item \textbf{Page Source:} The HTML and Document Object Model (DOM) structure that constitutes the web page. This includes critical elements such as forms, buttons, and interactive components that define the layout and available functionalities.
    \item \textbf{Metadata:} Ancillary data related to the current web session, including HTTP headers, cookies, and session-specific variables. This metadata provides additional context regarding the state of the application, reflecting conditions like user authentication, session persistence, or dynamic content adjustments.
\end{itemize}

State transitions occur when users interact with the application, such as through navigation, form submission, or button clicks. These transitions, captured as edges in the graph, form the relationships between states and drive the knowledge representation.

In complex web applications, multiple states may correspond to a single URL, but may vary due to session-specific factors or dynamically rendered content. For instance, a checkout page may present distinct states depending on whether items have been added to the cart or whether delivery options are available for the user’s location. These variations are captured through the combination of structural data and metadata.
\subsection{Action}
\label{sect:action}

In the context of this system, an \textit{Action} refers to a user-initiated operation or event that transitions the web application from one state to another. Actions represent the interaction points between the user and the web application, such as clicking a button, submitting a form, navigating to a new page, or triggering an AJAX request. These actions are fundamental to the system’s ability to explore and infer functionalities within the web application.

Actions are captured and represented as edges in the knowledge graph, where each edge connects two nodes (states) and denotes the transition caused by a specific interaction. The goal of the system is to not only capture the actions that lead to state transitions but also to rank and prioritize them based on their significance to the web application’s functionality.

Key characteristics of an action include:
\begin{itemize}
    \item \textbf{Action Type:} Actions can vary widely, from simple navigation (e.g., following a hyperlink) to complex interactions (e.g., filling out and submitting a form). These actions are categorized into types based on the nature of the interaction, such as clicks, form submissions, keyboard inputs, or dynamic event triggers (e.g., JavaScript events).

    \item \textbf{Action Context:} Each action is tied to a specific element in the DOM structure, such as a button, link, or form field. The context includes metadata such as the element’s attributes (e.g., ID, class) and its location within the page hierarchy. This context helps the system understand how the action relates to the structure of the web application.

    \item \textbf{Effect on State:} Actions are only significant if they result in a state change, meaning they transition the web application from one distinct state to another. The Functionality Inferring Module analyzes the effect of each action on the state, ensuring that only meaningful transitions are captured. For example, submitting a form might transition the user from a login page to a dashboard, whereas clicking a non-interactive element would not result in a state change.

    \item \textbf{Action Priority:} Not all actions contribute equally to the exploration of the application’s functionality. The system prioritizes actions that lead to new or unexplored states. Actions that produce trivial or redundant transitions (e.g., right-clicking or hovering over an element without causing a meaningful change) are deprioritized by the Re-ranking Module, ensuring that the exploration process is efficient.
\end{itemize}

\subsection{Functionality Inferring Module}
\label{sect:functionality-inferring-module}

The \textit{Functionality Inferring Module} is responsible for analyzing the current state of the web application and predicting potential actions that could transition the application to new states. This module synthesizes information from the current observation, previously explored functionalities, and the state-action history to identify and rank possible actions. The primary goal is to maximize the discovery of new functionalities and ensure that the system explores meaningful user interactions.

\begin{figure*}[tb]
	\begin{centering}
	\includegraphics[width=0.8\textwidth]{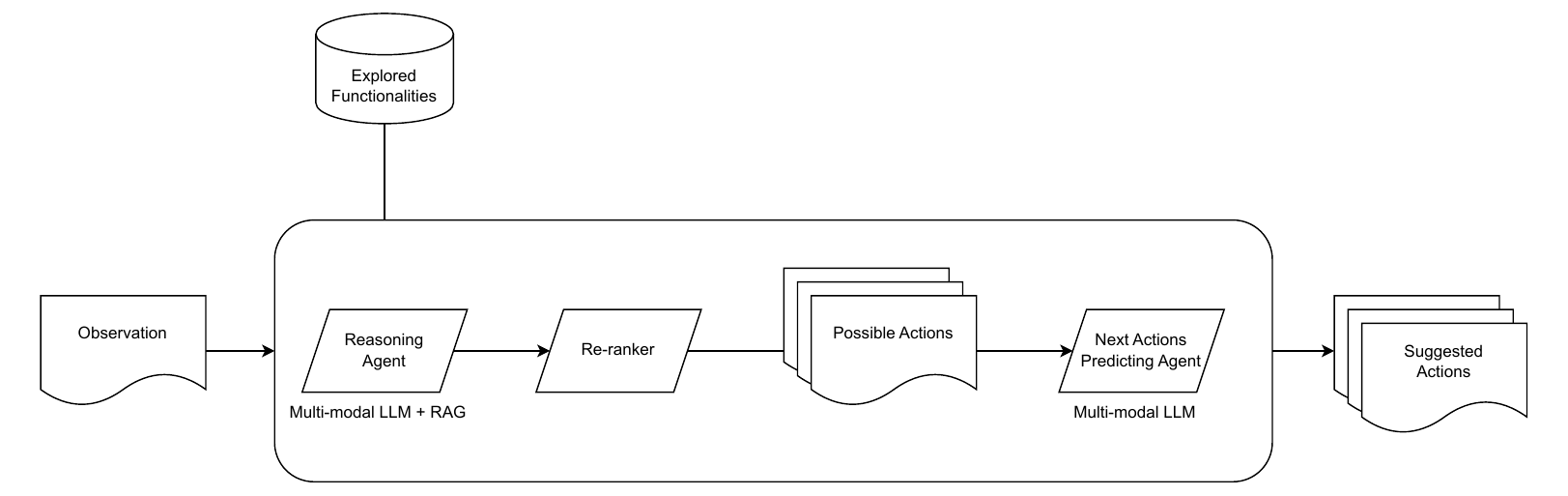}
	\caption{Functionality Inferring Module}
	\label{fig:figure-2}
	\end{centering}
\end{figure*}

The module comprises four key components:

\subsubsection{Reasoning Agent}
This agent processes the current observation—comprising the page source, screenshot, and metadata—alongside the record of previously explored functionalities. It synthesizes multiple queries to interface with the database, determining what functionalities have already been explored and what actions are possible given the current state and past interactions. The Reasoning Agent outputs a list of possible actions that can be performed based on the current and previous states, ensuring a thorough exploration of the application's functionalities.

\paragraph{Multi-modal LLM for State Understanding}
The Reasoning Agent employs a multi-modal LLM to comprehensively understand the current state of the web application. By analyzing various inputs—including the page source, screenshots, and session metadata—the LLM can generate a semantic and structural understanding of the application's current state.

\paragraph{Database Interface for Explored Functionalities}
In addition to understanding the current state, the Reasoning Agent interfaces with a database of previously explored functionalities. This ensures that the agent avoids redundant actions and focuses on unexplored areas of the application. The database stores all previously visited states and actions taken, forming a history of interactions with the web application. By querying this database, the agent can identify which actions have already been executed and which states have already been visited, allowing it to prioritize new interactions.

\paragraph{Generation of Possible Actions}
Based on the current state and the record of explored functionalities, the Reasoning Agent outputs a list of plausible actions that can be performed. These actions may include navigation, form submissions, button clicks, or more complex interactions involving multi-step processes.

\subsubsection{Re-ranking Module}
Once the Reasoning Agent generates a list of possible actions, the Re-ranking Module evaluates these actions and reorders them based on metrics such as entropy and expected reward. The objective is to prioritize actions that are most likely to uncover new functionalities or lead to significant state transitions, while deprioritizing trivial or redundant interactions (e.g., non-functional actions like right-clicking on an element). This dynamic re-ranking process ensures that the exploration of the application remains focused on discovering meaningful user flows and interactions, ultimately maximizing the system’s efficiency and effectiveness in navigating complex web applications.

\subsubsection{Next Actions Prediction Agent}
The Next Actions Prediction Agent uses a finetuned multi-modal LLM that refines the list of actions. It selects the top-ranked actions from the Possible Actions List and predicts the next best steps to take. This agent combines insights from both the re-ranked list and the system’s understanding of the web application to choose actions that will maximize the system’s overall reward.

The Functionality Inferring Module operates through a feedback loop. After executing each action, the module updates its understanding of the web application’s behavior, incorporating newly observed states and actions. This feedback ensures that the system continually improves its predictions and focuses on uncovering the most important functionalities.
\subsection{Action Executor}
\label{sect:action-executor}

The \textit{Action Executor} is responsible for executing actions or sequences of actions within the web application based on the inputs from the Re-ranking Module and Next Actions Prediction Agent. It performs interactions such as clicks, form submissions, and complex multi-step operations.

\subsubsection{Action Execution}
The Executor applies selected actions, which may involve single interactions (e.g., clicking a button) or multi-step sequences (e.g., form submissions). It handles various action types, including user interface actions, navigational transitions, and event-driven triggers like JavaScript.

\subsubsection{State Validation}
After executing actions, the Executor verifies whether the action resulted in a meaningful state change by capturing the updated page source, screenshots, and metadata. This validation is critical for updating the knowledge graph.

\subsubsection{Error Handling and Recovery}
In the case of failed actions due to issues like incorrect inputs or unhandled edge cases, the Executor logs the error and retries or performs recovery actions to restore the application to a stable state.
\subsection{Reward/Penalty Model}
\label{sect:reward_penalty_model}

The \textit{Reward/Penalty Model} quantifies the system’s progress in exploring meaningful functionalities within the web application. After the \textit{Action Executor} performs an action and returns the new state, this model evaluates the outcome by assigning a score between -1 and +1, indicating the action's value. Positive scores reflect significant progress, while negative scores highlight trivial or redundant actions.

\subsubsection{Rewarding Significant Progress}
Actions that lead to new state transitions or the discovery of unexplored functionalities receive positive rewards (closer to +1). For example, navigating from the home page to product listings or accessing the checkout process are high-reward actions as they reveal critical application behaviors.

\subsubsection{Penalizing Redundant Actions}
When actions result in trivial transitions (e.g., reaching a leaf node where no further meaningful actions can be taken), the model assigns penalties (closer to -1). This prevents the system from getting stuck in unproductive states, such as a "Thank you" page after purchase completion.

\subsubsection{Stopping Exploration}
If no actions produce a positive reward, the system halts exploration for that path. This ensures resources are not wasted on dead ends and exploration focuses on uncovering valuable transitions.

\subsubsection{Retrials}
In cases where the action taken results in a reward score close to the defined threshold but not sufficiently positive, this model will initiate a retrial if the maximum number of retries has not been exhausted.
\section{Experiments}
\label{sect:experiments}

To evaluate the effectiveness of our proposed approach, we conducted experiments on a real-world e-commerce website, \textit{Dentomart.com}, which specializes in selling dentistry equipment. This website includes a wide range of dynamic and interactive components, such as product search, filtering, cart management, and user authentication, making it an ideal candidate for testing the limitations of traditional parsers and the capabilities of our solution.

\begin{figure*}[tb]
    \centering
    \begin{subfigure}[b]{1\textwidth}
        \centering
        \includegraphics[width=0.8\textwidth]{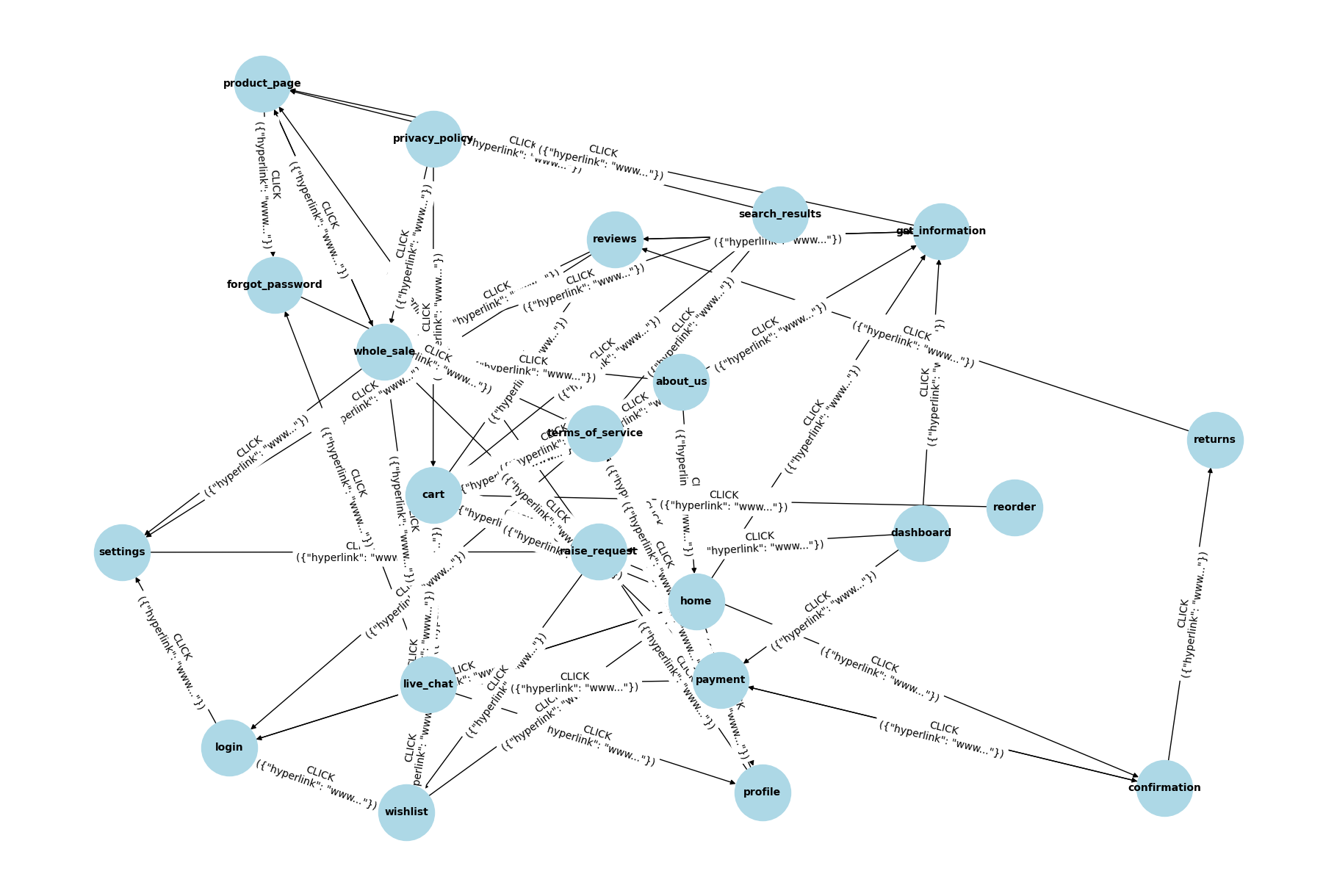} 
        \caption{A simplified and incomplete graph generated by traditional parsers, showcasing limited user interactions with only basic state transitions. This representation is insufficient for capturing the complexity of dynamic web applications.}
        \label{figure:figure-3a}
    \end{subfigure}
    \hfill
    \begin{subfigure}[b]{1\textwidth}
        \centering
        \includegraphics[width=0.8\textwidth]{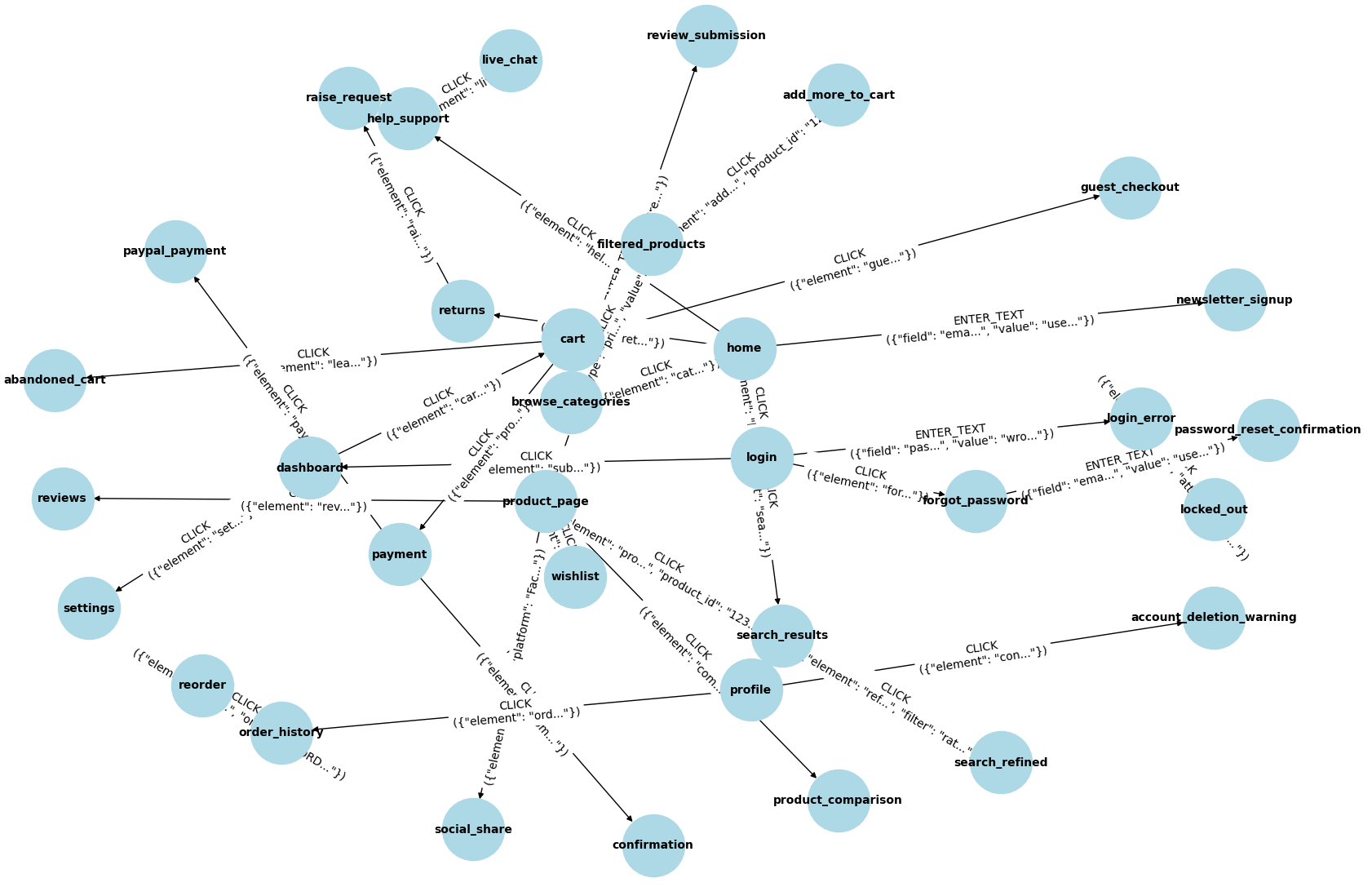} 
        \caption{A detailed graph generated by our proposed solution, illustrating a more comprehensive representation of user interactions with dynamic state transitions, complex user flows, and action-based edges that offer richer insights into web application behavior.}
        \label{figure:figure-3b}
    \end{subfigure}

    \caption{Comparison of traditional parsers (Figure 3a) vs. our proposed solution (Figure 3b) for modeling web application behavior}
    \label{figure:figure-3}
\end{figure*}

\subsection{Traditional Parser Setup}

For comparison purposes, we implemented a traditional parser using the \textit{Scrapy} framework, which is commonly used for web scraping. Our Scrapy-based parser follows a basic depth-first approach for crawling web pages, extracting hyperlinks, and collecting static page content. Specifically, the parser was configured with the following parameters:
\begin{itemize}
    \item \textbf{Max crawl depth}: 3 levels
    \item \textbf{Follow redirects}: Enabled
    \item \textbf{Concurrent requests}: 8
    \item \textbf{User-agent rotation}: Implemented to mimic various browsers
\end{itemize}

\subsection{Proposed Solution Setup}

To address the limitations of traditional parsers, we build a directed graph that models Dentomart.com. The graph was constructed using the following hyperparameters:
\begin{itemize}
    \item \textbf{min\_reward}: 0 — The minimum threshold for reward-based transitions, used to eliminate low-value or redundant actions.
    \item \textbf{max\_leaf\_branches}: 999 — The maximum number of branches a leaf node can have before being pruned.
    \item \textbf{max\_consecutive\_actions}: 5 — The maximum number of consecutive actions allowed within a single state before forcing a transition.
    \item \textbf{max\_retries}: 3 — The maximum number of attempts for each action before considering it a failed interaction.
\end{itemize}

\subsection{Evaluation Metrics}

We compared the performance of the traditional Scrapy-based parser and our proposed solution using the following key metrics:

\subsubsection{State coverage}
The number of unique states visited by each method. For traditional parsers, a unique state is typically identified by a unique URL within the domain. Higher state coverage is better because it reflects a more comprehensive exploration of the web application, including all key functionalities and dynamic states.

\subsubsection{Edge complexity}
The total number of edges (interactions) captured between states. Ideally, this value is close to $n-1$, where $n$ is the total number of states. This indicates minimal distractions between flows, with no unrelated or redundant transitions, suggesting that each transition contributes meaningfully to the functionality being explored.

\subsubsection{Failure recovery}
The ratio of actions that failed on the first attempt but succeeded within the \textit{max\_retries} allowed by the system. A higher value indicates better robustness, as the system is able to recover from failures and explore alternative paths or retry actions successfully.

\subsubsection{Time to completion}
The total time taken by each method to complete the crawl. In this metric, lower values are better, as faster completion means more efficient exploration.

\subsubsection{Graph density}
This metric measures the ratio of actual edges to the total possible edges in the graph. Lower density implies that the graph is not overly crowded with meaningless connections, which would indicate more structure and clarity in the transitions between states.

\subsubsection{Shortest path length}
The average shortest path length between any two nodes in the graph, which measures the overall connectivity. A longer shortest path may indicate more unique states and deeper exploration of the application. For traditional parsers, the path length may be shorter due to fewer unique states being captured, while in our approach, it is likely longer because of the broader state coverage and complex interactions.

\subsubsection{Betweenness centrality}
This metric measures the importance of nodes in connecting different parts of the graph. A higher value suggests that certain states (nodes) serve as crucial junctions in the web application’s flows. This can be useful in identifying critical pages, such as login screens or checkout processes, that play a significant role in navigating through the application. A higher betweenness centrality is often desirable for identifying key interaction points in the user journey.

\section{Results}
\label{sect:results}

We conducted experiments on the Dentomart web application, comparing a traditional Scrapy-based parser with our proposed solution. The traditional parser captured basic static states and edges but struggled with dynamic content and failed to capture user-triggered behaviors like form submissions or AJAX-based content loading. In contrast, our proposed solution effectively modeled dynamic state transitions, capturing significantly more states and edges, though with a longer time to completion.

\subsection{Key Metrics}

Table \ref{table:results-comparison} summarizes the key differences between the traditional parser and our solution.

\begin{table}[h!]
    \centering
    \resizebox{\columnwidth}{!}{%
    \begin{tabular}{|l|c|c|}
        \hline
        \textbf{Metric} & \textbf{Traditional Parser} & \textbf{Proposed Solution} \\
        \hline
        State complexity (no. of states) & 24 & \textbf{95} \\
        \hline
        Edge complexity (no. of edges) & 86 & \textbf{94} \\
        \hline
        Failure recovery rate & N/A & \textbf{0.72} \\
        \hline
        Time to completion (seconds) & \textbf{300} & 5500 \\
        \hline
        Graph density & 0.72 & \textbf{0.15} \\
        \hline
        Shortest path length & 2.1 & \textbf{6.4} \\
        \hline
        Betweenness centrality (avg) & \textbf{0.59} & 0.02 \\
        \hline
    \end{tabular}
    }
    \caption{Comparison of traditional parser vs. proposed solution across key metrics.}
    \label{table:results-comparison}
\end{table}

The proposed solution achieved much better \textbf{state coverage} and \textbf{edge complexity}, capturing a significantly higher number of interactions. It also excelled in detecting dynamic behaviors with robust \textbf{failure recovery}, which the traditional parser missed entirely. Notably, the average \textbf{betweenness centrality} in our approach is lower than in traditional parsers. This is primarily due to the higher number of unique states our model identifies, each representing distinct state transitions triggered by user interactions and dynamic content. These unique states are less likely to serve as intermediary nodes across multiple flows, reducing their overall centrality in the graph. In contrast, traditional parsers often rely on hyperlinks, leading to more shared or reused states across different flows, which increases the likelihood of those states acting as bridges. Consequently, while traditional parsers produce more centralized graphs, our method results in a more diverse and decentralized structure with fewer critical intersections.

\subsection{Procedurally Generated Test Cases by Graph Traversal}

One significant downstream application of the rich graph-based representation produced by our solution is automated testing. Each root-to-leaf path in the graph corresponds to a unique user flow or interaction sequence within the web application. Since every functionality within the application is reflected as a distinct path, we were able to procedurally generate test cases by traversing these root-to-leaf paths.

In total, the proposed solution generated \textbf{51 unique test cases}, covering a wide range of user interactions such as logging in, searching for products, adding items to the cart, and completing the checkout process. By automating the traversal of these paths, test cases can be programmatically generated, ensuring comprehensive coverage of user interactions, including edge cases and complex workflows that may be difficult to test manually.

This procedural test generation ensures that the entire functionality of the application is tested, and the dynamic states and behaviors captured by the graph provide detailed insight into every possible user interaction. As a result, the system can easily identify potential bugs, usability issues, or performance bottlenecks by systematically exploring each unique path through the application.


\end{document}